\documentclass[]{spie}  

\usepackage{amsmath,amsfonts,amssymb}
\usepackage{graphicx}
\usepackage[colorlinks=true, allcolors=blue]{hyperref}

\title{Design and development of the SOXS calibration unit}

\author[a,b]{Hanindyo Kuncarayakti}
\author[c]{Jani Achr\'en}
\author[d]{Sergio	 Campana}
\author[e]{Riccardo	Claudi}
\author[f]{Pietro	Schipani}
\author[d]{Matteo Aliverti}
\author[e]{Andrea Baruffolo}
\author[g,h]{Sagi Ben-Ami}
\author[e]{Federico Biondi}
\author[f]{Giulio Capasso}
\author[i,j]{Rosario Cosentino}
\author[k]{Francesco D'Alessio}
\author[d]{Paolo	D'Avanzo}
\author[g]{Ofir Hershko}
\author[d]{Marco Landoni}
\author[j]{Matteo Munari}
\author[l,m]{Giuliano Pignata}
\author[n]{Adam Rubin}
\author[o,j]{Salvatore Scuderi}
\author[k]{Fabrizio Vitali}
\author[p]{David	Young}
\author[q,m]{Jos\'e Antonio Araiza-Duran}
\author[r]{Iair Arcavi}
\author[l,n]{Anna Brucalassi}
\author[g]{Rachel Bruch}
\author[e]{Enrico Cappellaro}
\author[f]{Mirko Colapietro}
\author[f]{Massimo Della Valle}
\author[e]{Marco De Pascale}
\author[j]{Rosario Di Benedetto}
\author[f]{Sergio D'Orsi}
\author[g]{Avishay Gal-Yam}
\author[d]{Matteo Genoni}
\author[i]{Marcos Hernandez}
\author[b,a]{Jari Kotilainen}
\author[s]{Gianluca Li Causi}
\author[a]{Seppo Mattila}
\author[e]{Kalyan Radhakrishnan}
\author[g]{Michael Rappaport}
\author[e]{Davide Ricci}
\author[d]{Marco Riva}
\author[e]{Bernardo Salasnich}
\author[p]{Stephen Smartt}
\author[j]{Ricardo Zanmar Sanchez}
\author[t]{Maximilian Stritzinger}
\author[i]{Hector Ventura}

\affil[a]{Tuorla Observatory, Dept. of Physics and Astronomy, FI-20014 University of Turku, Finland}
\affil[b]{Finnish Centre for Astronomy with ESO (FINCA), FI-20014 University of Turku, Finland}
\affil[c]{Incident Angle Oy, Capsiankatu 4 A 29, FI-20320 Turku, Finland}
\affil[d]{INAF - Osservatorio Astronomico di Brera, Via Bianchi 46, I-23807 Merate (LC), Italy}
\affil[e]{INAF - Osservatorio Astronomico di Padova, Vicolo dell'Osservatorio 5, I-35122 Padova, Italy}
\affil[f]{INAF - Osservatorio Astronomico di Capodimonte, Salita Moiariello 16, I-80131 Napoli, Italy}
\affil[g]{Weizmann Institute of Science, Herzl St 234, Rehovot, 7610001, Israel}
\affil[h]{Harvard-Smithsonian Center for Astrophysics, Cambridge, USA}
\affil[i]{INAF - Fundaci\'on Galileo Galilei, Rambla J.A. Fern\'andez P\'erez 7, E-38712 Bre\~na Baja (TF), Spain}
\affil[j]{INAF - Osservatorio Astrofisico di Catania, Via S. Sofia 78, I-95123 Catania, Italy}
\affil[k]{INAF - Osservatorio Astronomico di Roma, Via Frascati 33, I-00078 Monte Porzio Catone, Rome, Italy}
\affil[l]{Universidad Andres Bello, Avda. Republica 252, Santiago, Chile}
\affil[m]{Millennium Institute of Astrophysics (MAS), Nuncio Monse\~nor Sotero Sanz 100, Providencia, Santiago, Chile}
\affil[n]{European Southern Observatory, Karl Schwarzschild Strasse 2, D-85748, Garching bei M\"unchen, Germany}
\affil[o]{INAF - Istituto di Astrofisica Spaziale e Fisica Cosmica, Via Corti 12, I-20133 Milano, Italy}
\affil[p]{Queen's University Belfast, Belfast, County Antrim, BT7 1NN, UK}
\affil[q]{Centro de Investigaciones en Optica A. C., Loma del Bosque 115, Lomas del Campestre, 37150 Leon Guanajuato, Mexico}
\affil[r]{Tel Aviv University, Department of Astrophysics, 69978 Tel Aviv, Israel}
\affil[s]{INAF - Instituto di Astrofisica e Planetologia Spaziali, Via Fosso del Cavaliere, I-00133 Roma, Italy}
\affil[t]{Aarhus University, Ny Munkegade 120, D-8000 Aarhus, Denmark}

\authorinfo{Further author information: (Send correspondence to H.K., E-mail:hankun@utu.fi)\\
}

\begin{document} 
\maketitle

\begin{abstract}
SOXS is a new spectrograph for the New Technology Telescope (NTT), optimized for transient and variable objects, covering a wide wavelength range from 350 to 2000 nm. SOXS is equipped with a calibration unit that will be used to remove the instrument signatures and to provide wavelength calibration to the data. The calibration unit will employ seven calibration lamps: a quartz-tungsten-halogen and a deuterium lamp for the flat-field correction, a ThAr lamp and four pencil-style rare-gas lamps for the wavelength calibration. The light from the calibration lamps is injected into the spectrograph mimicking the f/11 input beam of the NTT, by using an integrating sphere and a custom doublet. The oversized illumination patch covers the length of the spectrograph slit homogeneously, with $<1\%$ variation. The optics also supports the second mode of the unit, the star-simulator mode that emulates a point source by utilizing a pinhole mask. Switching between the direct illumination and pinhole modes is performed by a linear stage. A safety interlock switches off the main power when the lamp box cover is removed, preventing accidental UV exposure to the service personnel. All power supplies and control modules are located in an electronic rack at a distance from the telescope platform. In this presentation we describe the optical, mechanical, and electrical designs of the SOXS calibration unit, and report the status of development in which the unit is currently in the test and verification stage.
\end{abstract}


\section{INTRODUCTION}
\label{sec:intro}  

The Son of X-Shooter (SOXS) is a new spectrograph for the ESO NTT telescope at La Silla observatory, Chile (see Schipani et al., this volume, also reference \citenum{schipani18}). It is designed for the observations of astrophysical transient and variable objects, allowing for a rapid spectrum acquisition and characterization throughout the near-ultraviolet and infrared wavelengths. The instrument is now currently under construction and scheduled to be on-sky in 2022.

SOXS is designed to cover a broad wavelength range from 350 to 2000 nm at the resolution of $R \approx 4500$, through simultaneous use of the ultraviolet-visual (UV-VIS) arm covering 300-850 nm and the near infrared (NIR) arm covering 800-2000 nm (Rubin et al., Vitali et al., this volume). The observational spectrum across the SOXS wavelength range will have to be corrected and calibrated, in order to produce science-ready data products. The final-reduced SOXS spectrum is expected to be free of instrumental signatures and calibrated in both the wavelength and flux axes. The calibration unit (CBX) subsystem of SOXS is thus designed to fulfill this purpose.

\section{DESIGN}

As in the case of SOXS, the design of the CBX is inspired by the X-Shooter\cite{vernet11} instrument at the ESO Very Large Telescope at Paranal Observatory, Chile. Following the X-Shooter design, there are 7 lamps used in SOXS CBX:
\begin{itemize}
\item Quartz-tungsten-halogen (QTH), for flatfielding (VIS and NIR): Osram 64425
\item Deuterium (D2), for flatfielding (UV): Hamamatsu L6301-50
\item Thorium-argon (ThAr), for wavelength calibration (UV-VIS): Photron hollow cathode
\item Four rare gas lamps of Ar, Hg, Ne, and Xe, for wavelength calibration (NIR): Newport pencil-style penray
\end{itemize}

\begin{figure} [t]
   \begin{center}
   \includegraphics[width=0.7\textwidth]{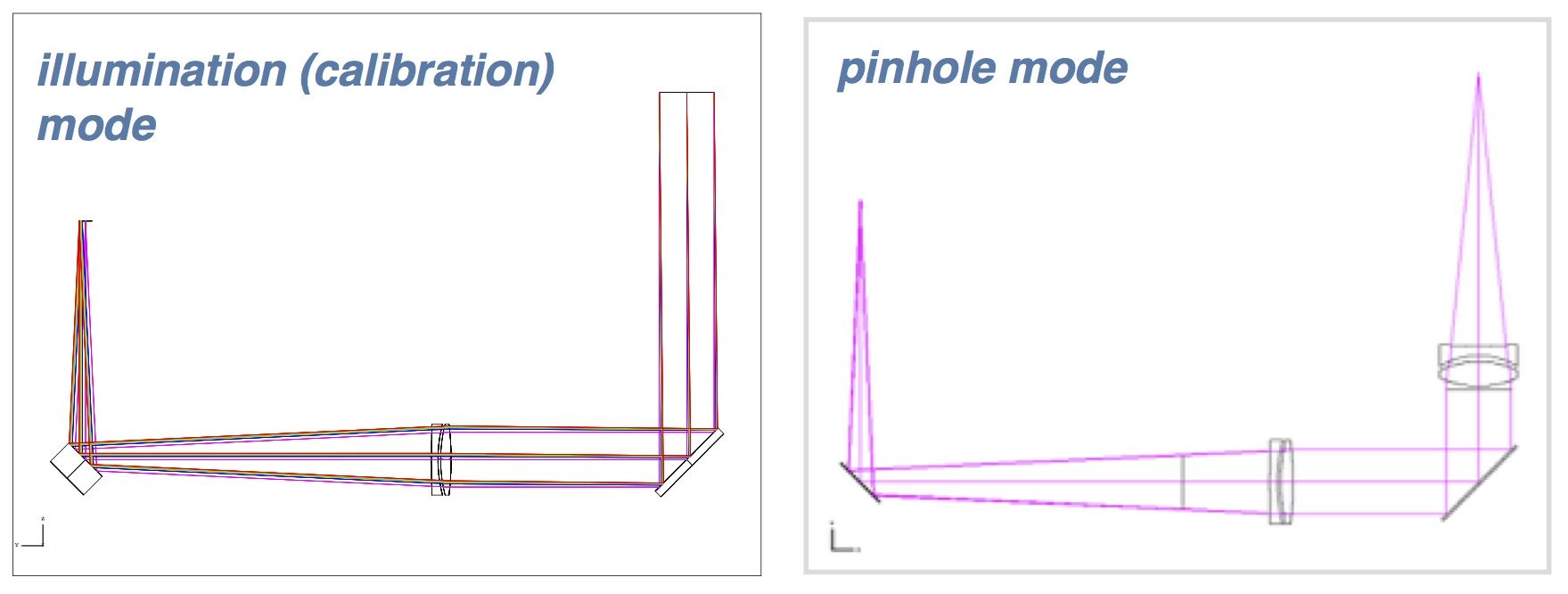}
   \end{center}
   \caption[example] 
   { \label{fig:modes} CBX optical design in two modes, illumination (left) and pinhole (right). The light coming from the integrating sphere at upper right is brought by the relay optics to the calibration selection mirror and instrument focal plane inside the common path on the left-hand side of the diagrams.}
   \end{figure}

\begin{figure} [b]
   \begin{center}
   \includegraphics[width=0.9\textwidth]{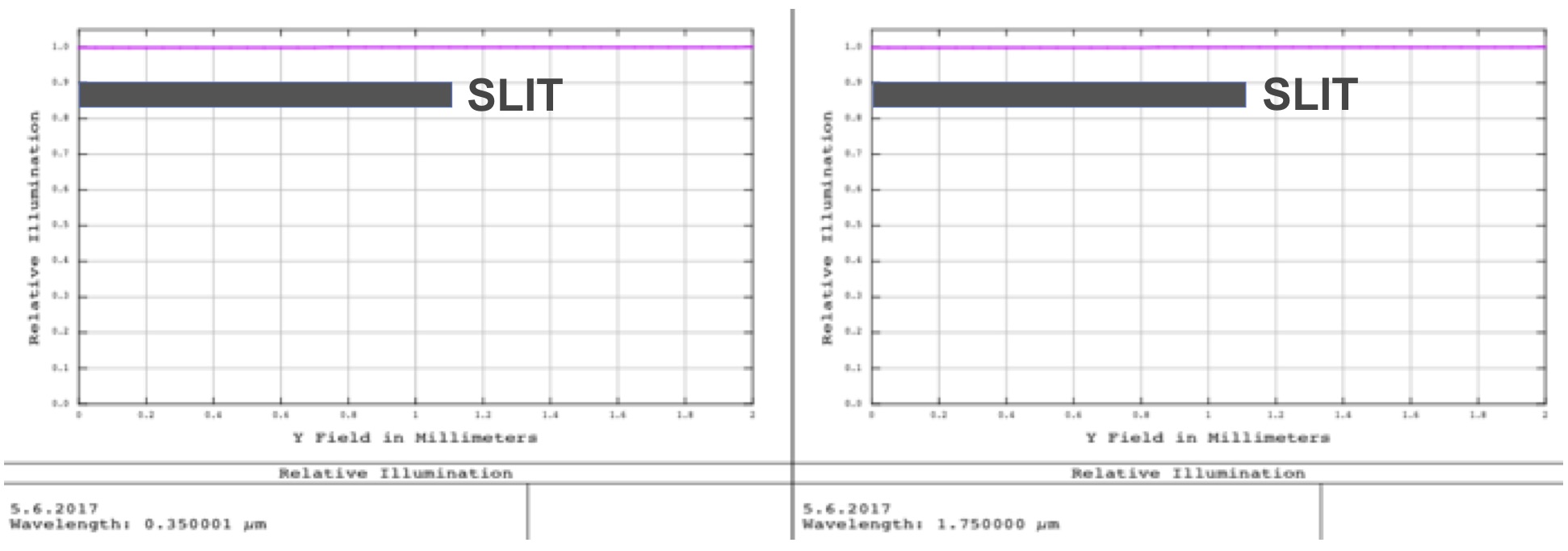}
   \end{center}
   \caption[example] 
   { \label{fig:relillum} 
Expected relative illumination of the CBX across the field of view, at 350 nm (left) and 1750 nm (right). Within the slit dimensions ($\pm$ 1.1 mm), the relative illumination (magenta lines) is homogeneous at the level of $1\%$.}
   \end{figure}

These types of lamps have been used as reliable calibration sources in a number of instruments at other large facilities\cite{kerber08a,kerber08b}.
The 7 lamps of the CBX feed the 4-port integrating sphere (Labsphere, Spectralon coated), with the configuration of 1 port holds the 4-penray cluster, 1 port holds the QTH lamp, 1 port is used together for the ThAr and D2 lamps, and 1 port is used for the exit beam. The penray cluster has individual tube holders for each penray, which allow for the adjustment of the exposed length of the penrays and in effect, balancing of the combined output spectrum from the 4 lamps.
The light exiting from the integrating sphere enters an aperture and is received by a 45$^{\circ}$ fold mirror that directs the light 90$^{\circ}$ to a doublet lens. The doublet focuses the beam towards the calibrator selector mirror inside the SOXS common path, and finally to the instrument focal plane and slit. This optical design is presented in Figure~\ref{fig:modes}, left panel. In this direct illumination mode, the CBX provides uniform illumination ($<1\%$ variation) across a $d = 4.8$ mm patch covering the whole 12 arcsec (2.2 mm at the focal plane) slit length (Figure~\ref{fig:relillum}) for the science calibrations. The CBX is designed to feed SOXS with an f/11 beam to simulate the NTT telescope beam.

\begin{figure} [t]
   \begin{center}
   \includegraphics[width=\textwidth]{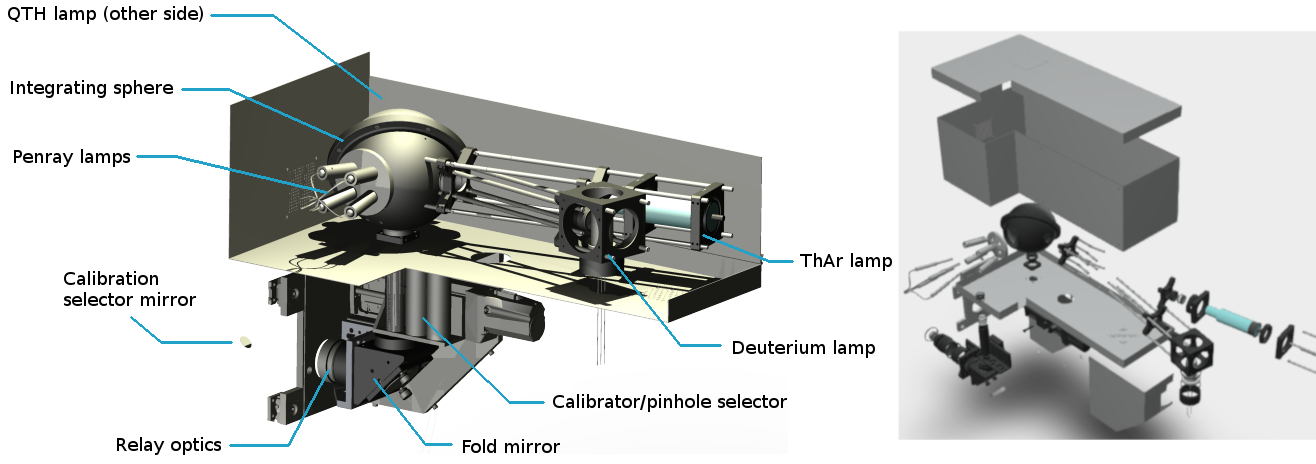}
   \end{center}
   \caption[example] 
   { \label{fig:cbxrender} 
Rendered image of the CBX, showing the components. An exploded image is also shown (right). }
   \end{figure}

   In addition to the direct illumination mode, the CBX is also equipped with a pinhole assembly to simulate a point source (Figure~\ref{fig:modes}, right panel). This pinhole assembly utilizes another doublet lens, and is positioned on a DC motorized linear stage Physik Instrumente PI L-406.40DD10. This allows for the selection between direct illumination and pinhole modes in the SOXS operations. Based on the optical design, the mechanical structures are designed to support the lamps and optics. A lightbox houses the integrating sphere and lamps, whereas the relay optics and linear stage are housed by the optics box underneath (Figure~\ref{fig:cbxrender}). The top lid of the lightbox is removable to enable lamp servicing when required. The lid is equipped with a sensor (Schmersal BNS 260-02Z-ST-L) which activates the Schmersal SRB301MC safety interlock circuit and cuts off the electricity to the lamps when the top lid is opened. This feature is used to prevent UV exposures to the service personnel when accessing the lamps accidentally without switching off the lamps first.

The power supplies of the calibration lamps are housed in a separate subrack. The CBX subrack will eventually be a part of the main SOXS electronics subrack, which is used to control the instrument through extension cables. In the case of CBX, the extension cables connect the lamps and their respective power supplies, and also the stage motor and the interlock sensor with the control modules. The control system is based on Beckhoff PLC. The lamp power supplies draw power from the mains 230 V AC input, whose Schuko lines are controlled on or off through ES2622 Beckhoff relay modules.  On the front panel of the subrack, hour meters are installed to monitor the lamp usage time and allow for scheduled maintenance. The connections between the lamps and respective power supplies are monitored by current transducers, which detect the current running when the lamp is on and sends signal to the lamp hour meters and Beckhoff ES1008 digital input module. It is therefore possible to monitor the status and remaining lifetime of the individual lamps. Two Kniel power supplies are installed in the subrack to provide 24 V DC lines to power the modules as well as the interlock, lamp current transducers and hour meters.

\section{DEVELOPMENT}

The design of the CBX was finalized in mid-2018. Following the Final Design Review, the procurement and manufacturing of the parts started. The construction of the subsystem was conducted in the premises of Department of Physics and Astronomy, University of Turku, in Turku, Finland.

    \begin{figure} [h!]
   \begin{center}
   \includegraphics[width=0.8\textwidth]{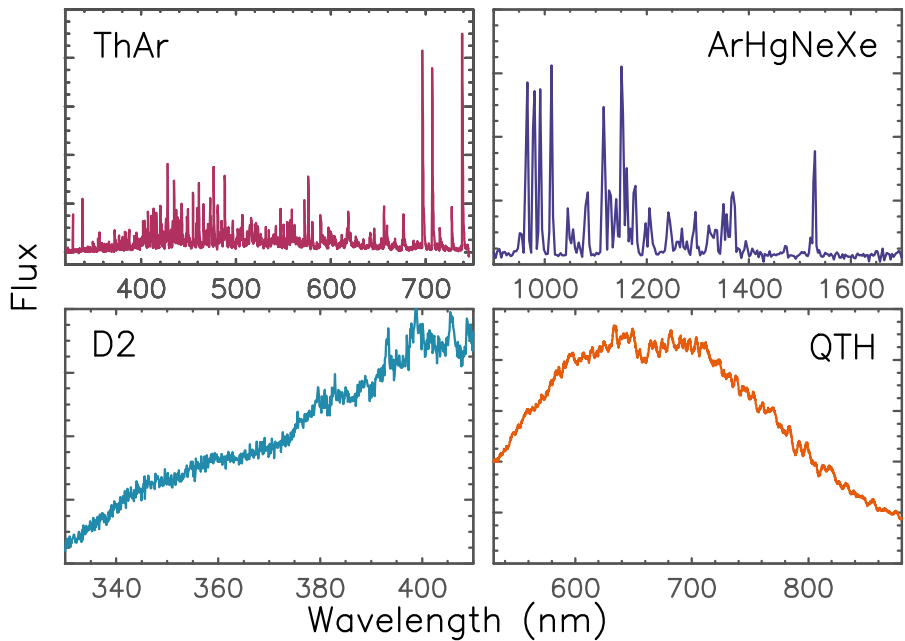}
   \end{center}
   \caption[example] 
   { \label{fig:cbxspec} 
Spectra of CBX calibration lamps as measured in the laboratory.}
   \end{figure}

Testing of the lamps spectral output was done in parallel with the construction and assembly of the mechanical parts. The lamp spectral output test is used to verify that the lamp is suitable for the intended use. The tests were done by feeding the light from the lamp to a fiber spectrometer, which were Thorlabs CCS100 (300-700 nm), Thorlabs CCS175 (500-1000 nm), and Hamamatsu C14486GA (950-1700 nm). Figure~\ref{fig:cbxspec} shows the examples of the measured spectra of the calibration lamps within their typical operation wavelength range.

       \begin{figure} [t]
   \begin{center}
   \includegraphics[width=0.5\textwidth,angle=180]{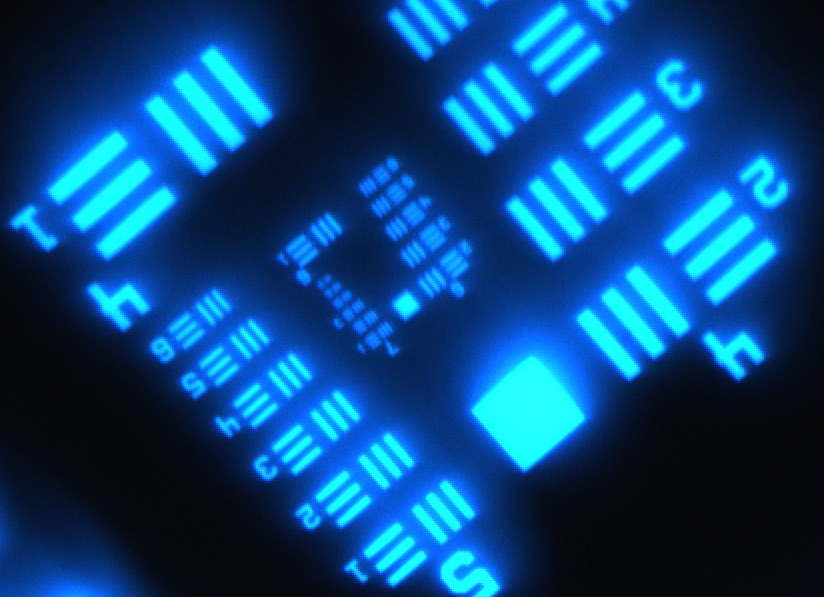}
   \end{center}
   \caption[example] 
   { \label{fig:usaf} 
USAF1951 resolution test for the CBX optics. The smallest resolved pattern is located in group 6, element 5 (corresponding to 4.9 $\mu$m resolution).}
   \end{figure} 
   
   The optics of the CBX was tested for the resolution and output f/\#. A light source was positioned at the focal point, and the ratio between the focal length  and the diameter of the full light beam at the plane of the system stop gives the f/\#. With this procedure the f/\# of the system was confirmed to be f/11 as designed. A resolution test was also done, using a USAF1951 resolution test target. The test result shows that the system resolution is around 5 $\mu$m (Figure~\ref{fig:usaf}).

       \begin{figure} [h!]
   \begin{center}
   \includegraphics[width=\textwidth]{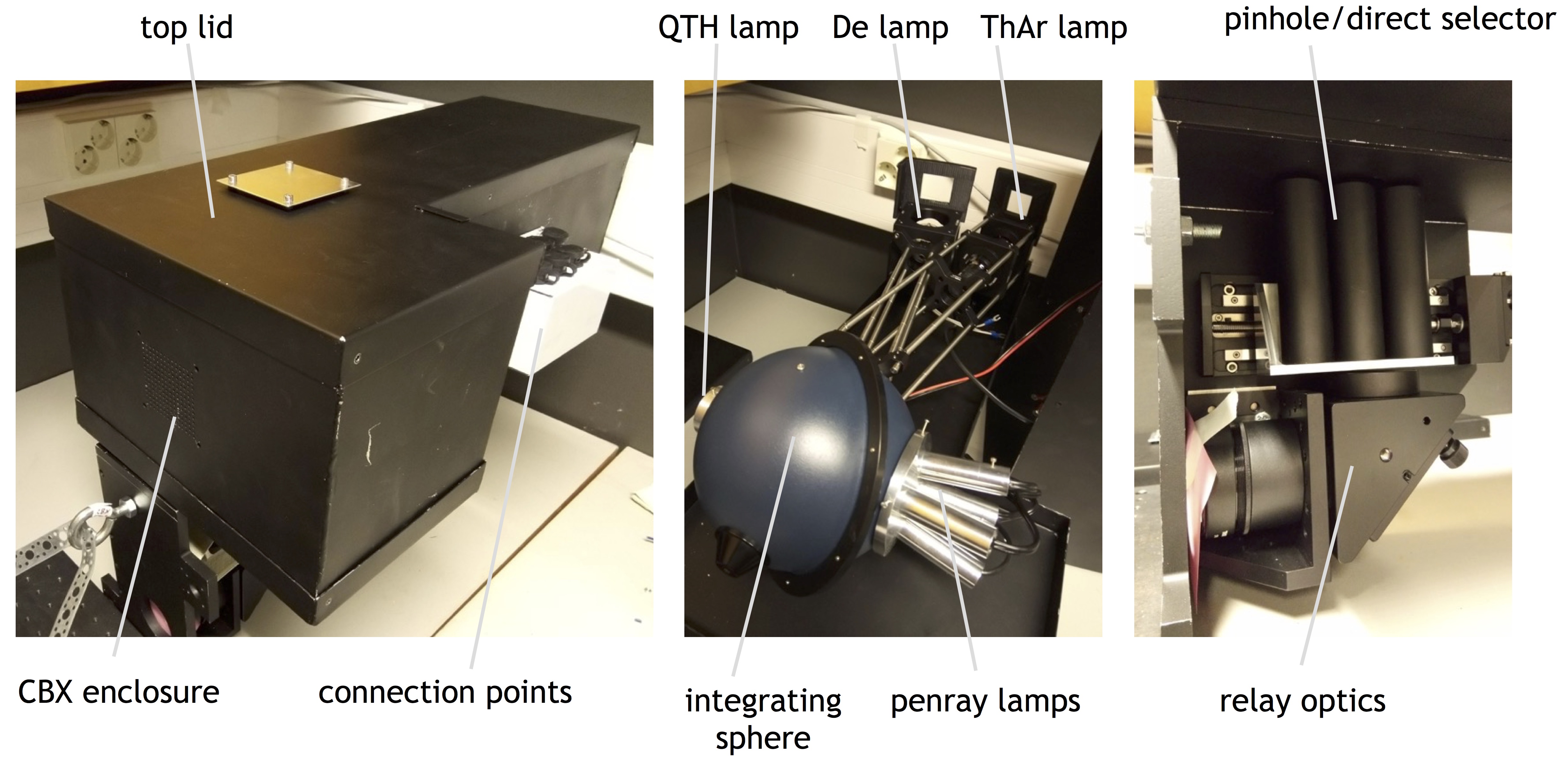}
   \end{center}
   \caption[example] 
   { \label{fig:cbxpic} 
Assembled CBX (left), with cover removed: light box (middle), and optics box (right). }
   \end{figure}

The CBX was assembled according to the mechanical design diagrams. Its total length is about 60 cm, height 55 cm, and thickness 30 cm. The mechanical parts are a combination of off-the-shelf components and self-machined parts done by the mechanical workshop of the University of Turku. The total weight of the subsystem is about 20 kg.
Figure~\ref{fig:cbxpic} shows the appearance of the CBX with and without covers of the light box and optics box. All the electrical connections in the CBX and subrack were designed to preserve the original lamp and power supply connectors, in order to facilitate replacements when necessary.

\begin{figure} [t]
   \begin{center}
   \includegraphics[width=0.75\textwidth]{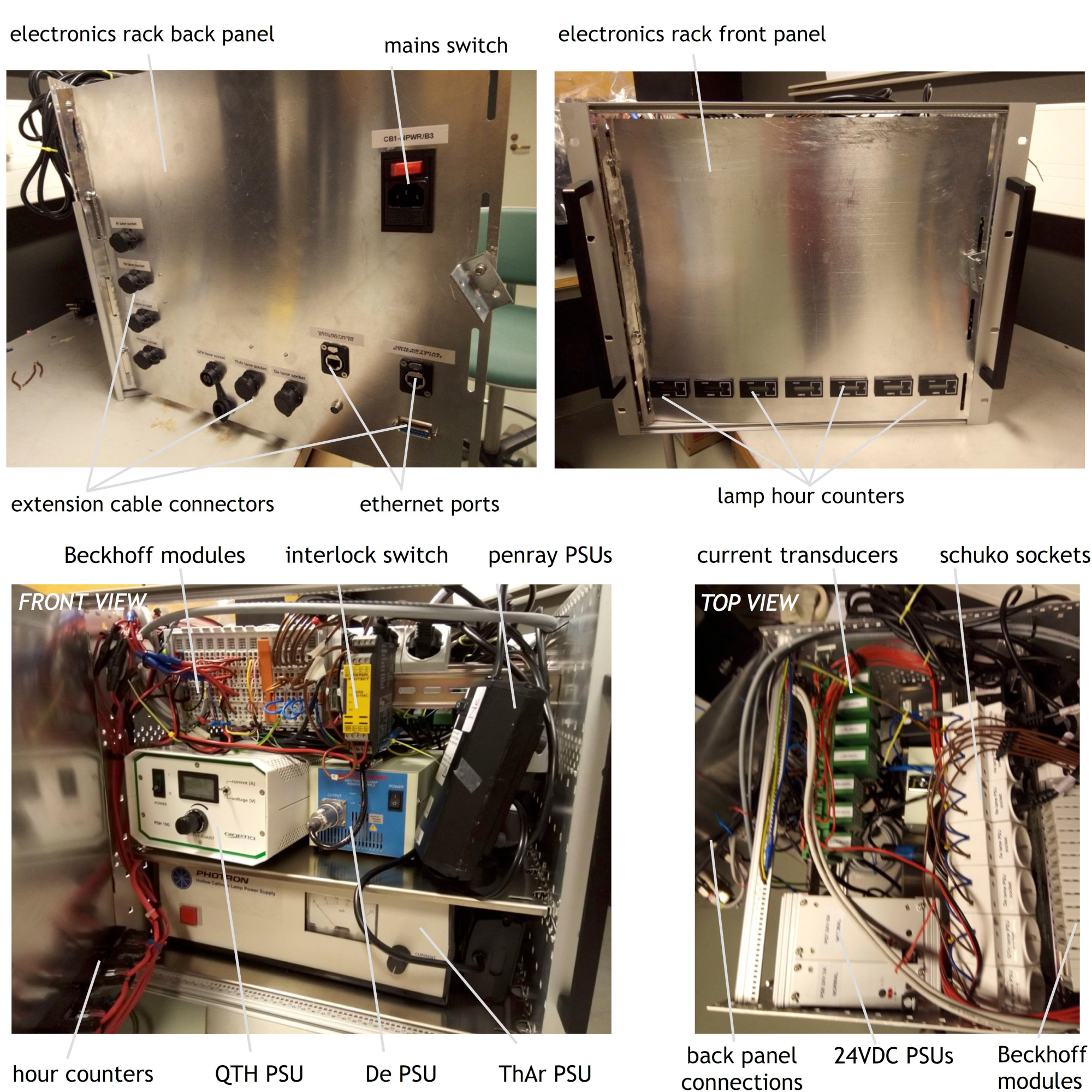}
   \end{center}
   \caption[example] 
   { \label{fig:rack} 
CBX electronics subrack, outside (upper panels) and inside (lower panels).}
   \end{figure}

   The construction of the CBX subrack was done together with the construction of the CBX itself. Figure~\ref{fig:rack} presents the electronics subrack of the CBX. The subrack occupies a 9U volume, and has two levels where the individual lamp power supplies are placed. The base level is occupied by the ThAr power supply, and the upper level is occupied by the D2 and QTH power supplies. The 4 penray power supplies are distributed as 3 on the upper level and 1 on the lower level. At the front portion of the subrack, the Beckhoff control modules are placed on a DIN rail, facing forward to facilitate wire management. On the same DIN rail are placed the Schmersal SRB301MC safety interlock guard door monitor and a switch that controls the mains current running into the ES2622 relay-controlled Schuko sockets on a second DIN rail. A third DIN rail is located at the rear side, and is used to attach the 7 current transducers.

Currently, the CBX and subrack are fully assembled and undergoing tests to verify the performance and functionalities. These tests include lamp spectral output, optics performance, stage movement, safety interlock function, and controls through software. The CBX is expected to be fully completed by the end of 2020 and integrated with the other SOXS subsystems in early 2021.

\section{SUMMARY}

The design and development status of the calibration unit (CBX) subsystem of the SOXS instrument is presented. The CBX uses 7 calibration lamps to perform flux and wavelength calibrations across the whole SOXS wavelength range of 350-2000 nm. The CBX is fully assembled and is currently undergoing verification tests, aimed for final completion by the end of 2020.

\acknowledgments 
 
We thank T. Kuusela for assistance with the optics.
We acknowledge support from the Academy of Finland through projects 324504, 328898, and 306531.

\bibliography{report} 
\bibliographystyle{spiebib} 

\end{document}